# Say It All: Feedback for Improving Non-Visual Presentation Accessibility

Yi-Hao Peng
yihaop@cs.cmu.edu
Carnegie Mellon University

JiWoong Jang
jiwoongj@cs.cmu.edu
Carnegie Mellon University

Jeffrey P. Bigham
jbigham@cs.cmu.edu
Carnegie Mellon University

Amy Pavel
apavel@cs.cmu.edu
Carnegie Mellon University

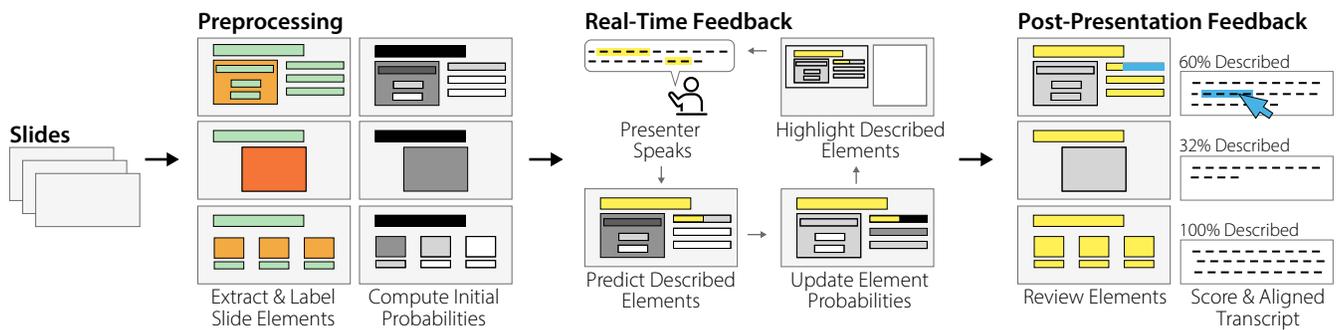

**Figure 1: Presentation A11y parses presentation slides and transcribes the presenter's speech in real-time to provide element-level feedback to presenters about whether they have verbally described visual content. Presenters can use real-time feedback to prompt them to speak about unaddressed slide elements or view post-presentation feedback to help them revise their slides.**

## ABSTRACT

Presenters commonly use slides as visual aids for informative talks. When presenters fail to verbally describe the content on their slides, blind and visually impaired audience members lose access to necessary content, making the presentation difficult to follow. Our analysis of 90 presentation videos revealed that 72% of 610 visual elements (*e.g.*, images, text) were insufficiently described. To help presenters create accessible presentations, we introduce Presentation A11y, a system that provides real-time and post-presentation accessibility feedback. Our system analyzes visual elements on the slide and the transcript of the verbal presentation to provide *element-level feedback* on what visual content needs to be further described or even removed. Presenters using our system with their own slide-based presentations described more of the content on their slides, and identified 3.26 times more accessibility problems to fix after the talk than when using a traditional slide-based presentation interface. Integrating accessibility feedback into content creation tools will improve the accessibility of informational content for all.

## 1 INTRODUCTION

Slide-based presentations are a prevalent medium for informative talks across education, business, and research. Presenters use visual content including text, images and videos on their slides to reinforce the concepts and structure of their verbal presentation. However, presenters often add visual content to their slides that they do not verbally describe, reducing the ability of audience members to understand the presentation. People who can see the presenter's slides pay less attention to the presenter's speech, and miss verbal information that is not visually present on the slide [35, 46]. People who cannot see the slides due to disability or situational impairment [47] miss information that the presenter conveys only visually.

General presentation guidelines [21, 32] and guidelines for creating accessible presentations [1, 6, 8, 22, 29, 31, 44] thus encourage presenters to achieve high correspondence between the spoken and visual content in their presentations by: minimizing the use of unnecessary visual content (*e.g.*, text, diagrams, videos), and verbally covering the text and media on their slides. Despite such guidelines, our formative analysis of 90 existing presentation videos (269 slides, and 610 visual elements) across several venues (*e.g.*, TED talks, seminars, and lectures) and domains (*e.g.*, selected topics in humanities, social sciences, and applied sciences) revealed that presenters fully described only 28% of slide elements. Further, 27% slide elements were fully absent from the verbal description, rendering them entirely inaccessible to blind and visually impaired audience members and others who can not see the slides.

When explicitly preparing to give an accessible presentation, applying high-level accessibility guidelines [22] to make low-level script and slide changes is difficult. Presenters without disabilities or experience giving accessible presentations may misapply high-level guidelines and overlook inaccessible slide elements. Practice talk audiences may not notice what slide elements have not been described, or lack time for low-level feedback. Adapting presentations to describe visual content at the time of the talk is even more challenging. On multiple occasions, the authors have observed presenters be asked to fully describe their slides for accessibility once their talk has begun. The typical presenter tries to do so, but quickly returns to not describing visual content fully as they attempt (and fail) to adjust their practiced speech on-the-fly, while respecting time limits, in the high-pressure situation of a public talk.

To help presenters make their presentations accessible, we introduce Presentation A11y (Figure 1). Presentation A11y helps presenters apply accessibility guidelines for live presentations [1, 6, 8, 22, 29, 31, 44] by providing content-specific *element-level feedback*. Presentation A11y first extracts, identifies, and labels visual elements on a presenter's slides (e.g., text, icons, images, and diagrams). As the presenter discusses a slide, Presentation A11y's *real-time feedback interface* transcribes the presenter's speech in real time and highlights what elements on the slide have been discussed on the presenter view. Afterwards, presenters can use the *post-presentation feedback interface* to review what slide elements lack description along with the speech that they used to describe each element. The real-time feedback can be used as a practice tool for rehearsed conference talks, during less scripted talks (*e.g.*, long classroom lectures, casual within-team presentations), and by expert presenters as a safeguard. The post-presentation feedback interface can be used with or without the real-time feedback display to improve future talk iterations.

We evaluated Presentation A11y in a study with 16 participants delivering their own slide presentations from a variety of venues (*e.g.*, conference, lecture) and domains (*e.g.*, writing, UI design). Using Presentation A11y's real-time feedback interface, participants described significantly more of the text and media content on their slides than they did without it. Using Presentation A11y's post-presentation feedback, presenters identified significantly more changes to make their presentation more accessible (*e.g.*, remove excess text) than they did without it (2.25 vs. 0.69 changes identified). All participants expressed enthusiasm about using Presentation A11y's feedback to improve their presentations in the future.

In summary, this paper contributes:

- An analysis of 90 existing slide presentations (269 slides) with respect to accessible presentation guidelines
- Presentation A11y, a tool for providing *element-level feedback* on correspondence between spoken and visual information
- A user study showing that with Presentation A11y presenters: (1) cover more slide content, and (2) identify more accessibility improvements than with guidelines alone

## 2 BACKGROUND

Our work relates to four key areas of prior work: (1) feedback for authoring accessible media, (2) systems for providing manual or automatic feedback on presentations, (3) algorithms for aligning recorded speech to slides, and (4) how visual and verbal slide-based presentations impact audience learning.

### 2.1 Feedback for Authoring Accessible Media

Prior work introduced systems to help people author accessible media according to guidelines, especially for accessible web applications [3, 19, 34, 39]. For instance, early work including Lift [28] and WAVE [19] helped people assess the level of their application's adherence to established web accessibility guidelines. To make the high-level guideline-based accessibility feedback easier to understand and apply, prior work provided visualizations to provide content-specific feedback: Takagi et al. [39] visualized the screen-reader reach time as a color gradient to help developers improve navigability, Sato et al. [34] visualized a "reading flow" path to help developers correct screen reader reading order, and Bigham et al. [3] visualized how screen reader users visited webpages to help developers recognize weak points. To help people create accessible presentations, we display high-level guidelines as content-specific feedback in the domain of live presentations by visualizing the coverage of slide elements in the spoken presentation.

### 2.2 Feedback for Authoring Presentations

Many people seek presentation feedback from peers [37] and audience members [12, 27]) to improve their presentations. Prior work helps presentation authors collect and organize audience feedback for future presentation improvement [12, 37]. Feedback from audience members can be invaluable, but audience member feedback is limited by audience expertise, time, multi-tasking ability, and willingness to participate. To help presenters independently improve their talks, prior work explored systems to automatically provide relevant presentation examples [45], in-situ prompts [2, 4], or performance feedback [9, 36, 40, 41]. Most of the prior work on presentation feedback focuses aspects of the presentation that are independent of slides such as vocal delivery (*e.g.*, volume modulation, use of filler words) and body language (*e.g.*, crossing arms) [36, 40, 45], but Trinh et al.'s system also provides users feedback on their level of success in covering their written script in their speech [41]. Microsoft's Presenter Coach [9] gives a single presentation-wide score for whether or not the speaker read off of their slides. Such systems all provide high-level feedback on delivery, but do not yet provide presenters *element-level feedback* to refine their narration based on their slides, and the slides of their presentation based on their narration.

Post-presentation feedback systems help presenters improve subsequent deliveries of the talk including subsequent practice talks or future lectures. But, in the case of less-rehearsed talks including lectures or casual presentations, presenters may opt for systems that provide real-time support for improving talk performance [2, 4]. SlidePacer by Brandão et al. [4] provides "pause" prompts to presenters based on the progress of the interpreter to make their presentations more accessible for hearing impaired audience members. We build on this work and guidelines for accessible presentations to help presenters make their slides more accessible for visually impaired audience members through real-time and post-presentation feedback.

## 2.3 Systems for Aligning Slides and Speech

Prior work considers how to align segments of a presentation video to presentation slides, for purposes such as note-taking and supporting slide-based navigation of a video recording [16–18, 38, 42, 43, 48]. Such prior work primarily considers: (1) aligning slides rather than individual elements to the presenters speech, and (2) recordings of the speech such that methods that align speech to slides can consider the content before and after the aligned word. Tsujimura et al. and Jung et al. consider real-time alignment of speech to slides for the purposes of students glancing at the slides to refer to the current speaker location [17, 42]. As we uniquely consider using speech to slide alignment to provide slide-element feedback in real-time, we match speech to text narrowly (e.g., using exact word matches rather than similarity-based matches) and we use a high-precision, low-recall approach (e.g., false negatives are more likely than false positives) to encourage further slide description and refinement.

## 2.4 Slide-Based Presentations and Learning

Our work helps presenters make their slide-based presentations accessible, but whether and how slides help audience members learn and retain the presentation content (or not) is a topic of ongoing research [21, 24, 25, 35, 46]. While slides can capture attention and facilitate learning by combining visual and verbal content, too much information can overload audiences' information processing resources [25]. When processing resources are overextended, audiences process verbal information better than they process visual information [24], such that mismatched verbal and visual slide content results in audiences retaining visual content over the verbal presentation [35, 46]. Presentation A11y thus encourages presenters to describe the content on their slides and remove content that they do not wish to describe to help assure that the loss of one channel (due to disability, situational impairment, or distraction) does not prevent audiences from gaining information. But, how often do presenters fail to match their visual and verbal content in practice? A prior analysis of slideshows (without their verbal presentations) found problems likely to cause temporal mismatches (e.g., ≥ 2 lines per bullet point, too many list items) [21]. To understand how well presenters describe their slides in-the-wild, we conduct an analysis of slides and their verbal presentations.

## 3 ACCESSIBLE PRESENTATION GUIDELINES

To help presenters make their presentations more accessible for audiences with a wide range of abilities, prior work provides high level guidelines for accessible presentations [1, 6, 8, 22, 29, 31, 44]. We leverage such guidelines, created by and in collaboration with disabled people to inform the design of our tool for presentation authors. Below, we summarize only guidelines for creating accessible live presentations rather than guidelines for making the presentation file itself more accessible (see guideline documents, WCAG [20], and prior work by Elias et. al. [7], for further guidance on distributing accessible slide files). We note that several guidelines for accessible live presentations align with prior research on learning from slide-based presentations (e.g., matching the visual and verbal content [35]), and guidelines for general presentations more broadly (e.g., simplicity, and structure [32]). We categorize key guidelines for live presentations in terms of three key accessibility considerations: visual (**VG**), auditory (**AG**) and cognitive (**CG**):

**Narrating visual content:** Verbally describe all pertinent visual information on the slides including text, images, graphics, and other visual aids (**VG1**). When describing visual content, use nouns instead of pronouns (**VG2**), and include context and regions of interest (**VG3**). Provide summaries of videos before playing them, and narrate the action in short phrases if possible (**VG4**).

**Language and pacing:** Use simple language, explaining jargon, acronyms and idioms (**CG1**). Allow sufficient time for the audience to process the presented information (**CG2**). Recognize and support delays that may exist with real-time captioning and sign language interpretation, and allow time for audience members reading captions to also read your slides (**AG1**).

**Positioning and sound:** Face the appropriate direction to support participants, sign language interpreters, lip readers and audio transcription in understanding your speech (**AG2**), and assure that all relevant sound loud enough to be audible from the sound system (**AG3**). Enable real-time transcription functionality (**AG4**).

**Slide design:** Use minimal visuals (**VG5**). To achieve minimal visuals, guidelines list rules of thumb such as no more than 5-7 lines/points and only about 5-6 words per line. Use proper font size, color, and typeface to make text readable (**VG6**), and use mixed case rather than all caps (**CG3**), for example a table of contents and progression cues. Provide structure cues (**CG4**), for example a table of contents and progression cues.

**Media alternatives:** Display equivalent text or graphics for pure audio media. Provide audio descriptions for videos (**VG7**); and if not, summarize then narrate the video as in VG4. Display equivalent text or graphics for pure audio media (**AG4**).

We build on such established high-level guidelines to provide presentation authors localized delivery and preparation feedback to make the guidelines concrete in the specific context of their own presentation. We focus on instantiating guidelines (**VG1-7**) to make presentations more non-visually accessible, but guidelines are interrelated such that simplifying slide content (**VG5**) can lead to less content on the slide for audience members to read during the presentation (**CG2, AG1**).

## 4 ANALYSIS OF PRESENTATION VIDEOS

To understand how well presenters describe the visual content on their slides, we conducted an analysis of verbal coverage of slide elements from existing slide-based presentation videos.

| Category | Slides | Elems | Total | Comp. | Text | Comp. | Media | Comp. |
|---|---|---|---|---|---|---|---|---|
| Humanities | 1.30 | 1.89 | 97 | 32% | 46 | 46% | 51 | 20% |
| Social Sci. | 1.27 | 2.32 | 112 | 26% | 63 | 41% | 49 | 6% |
| Natural Sci. | 1.03 | 2.50 | 109 | 22% | 57 | 33% | 52 | 10% |
| Applied Sci. | 0.86 | 2.80 | 102 | 27% | 69 | 38% | 33 | 6% |
| Formal Sci. | 1.30 | 2.52 | 150 | 33% | 110 | 44% | 40 | 5% |
| TED | 1.62 | 1.51 | 147 | 25% | 61 | 52% | 86 | 6% |
| Seminars | 1.05 | 2.96 | 222 | 27% | 145 | 34% | 77 | 12% |
| Classes | 0.79 | 2.74 | 201 | 33% | 139 | 42% | 62 | 13% |

**Table 1: For each field, the number of Slides per minute (Slides), the number of elements per slide (Elems), the total number of elements (Total), the % elements complete (Comp.), the number of text elements (Text) and % text elements complete, and the number of media elements (Media) and media elements complete (Comp.).**

## 4.1 Materials

To gain a broad view of presenter strategies when using slide-based visual aids in their presentations, we selected videos to represent a range of production and preparation levels (*e.g.*, TED talks, seminars, and lectures) and domains (*e.g.*, applied sciences, formal sciences, natural sciences, social sciences and humanities). For each of the 15 setting and domain type combinations (3 settings, 5 domain types), we selected a set of 6 videos of slide-based presentations, for a total of 90 slide-based presentation videos. To select the 6 videos for each domain/setting pair, we randomly sorted a list of academic disciplines (Wikipedia's Outline of Academic Disciplines [1]) then searched YouTube for "<academic discipline> <setting>" (*e.g.*, fluid mechanics lecture, marketing seminar). If the search was too narrow to return relevant results, we ascended the academic discipline hierarchy by one (*e.g.*, "Helminthology TED talk" –> "Zoology TED talk"). We then selected the first lecture that included slides as a visual aid (based on the thumbnail). To select a short segment from each presentation video, we randomly selected a starting point (at least 2 minutes from the start and 1 minute from the end) then scrubbed before that point to find the starting point of a slide and created a segment from there that included at least 2 slides and their explanations, and was at least 2 minutes in duration.

## 4.2 Analysis

For each video segment, we labelled each slide element with: (1) a slide element type label (*e.g.*, text-title, graph, image, diagram, video), (2) a coverage amount label (*e.g.*, none, little, half, most, complete), and (3) explicit reference label to address whether or not the presenter explicitly referred to visual content on the slide (*e.g.*, "as you can see here"). Two annotators (authors of this paper, sighted) coded 10 randomly selected video clips together to establish definitions for the coverage levels (similar to prior work [13]) and type labels by considering the adherence to delivery-related guidelines VG1-4, and VG7, then independently coded a 20% sample of the videos and achieved substantial agreement on the first pass across label types ($\kappa$=0.65-0.90, Landis and Koch [23]). The annotators then applied the established codes to the remaining videos independently.

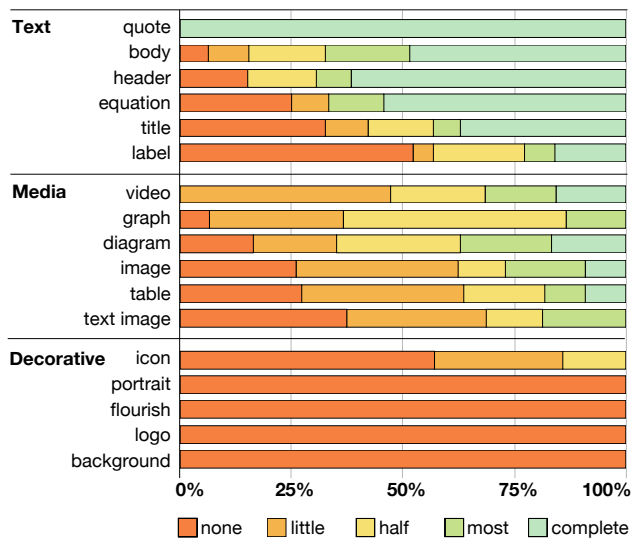

**Figure 2: The distribution of description levels for different types of slide elements including text elements (e.g., quote, body, header), image elements (e.g., video, graph, diagram), and decorative elements (e.g., icon, portrait, background).**

## 4.3 Results

Slide-based presentation video clips in our dataset featured 1.15 ($\sigma$ = 0.76) slides per minute and 2.40 ($\sigma$ = 0.93) elements per slide, with TED videos and humanities videos containing both the fewest elements per slide and the highest number of slides per minute (Table 1). Overall, presenters often did not describe all of the informative elements on the slide with 72% of non-decorative slide elements missing a description for at least some key information. The presenter's verbal coverage of slide elements varied by element type (Figure 2).

**Describing text:** Across all domains and venues, speakers covered text elements in speech more thoroughly than they did media elements, with TED videos and humanities talks containing higher completion levels for text (52% and 46% respectively) than other domains (Figure 1). Quotes, headers, and equations were the most likely to be completely described of the text elements (Figure 2). While quotes and equations often are described verbatim (Figure 3A), headers typically indicated important slide-specific topics that were often covered in the course of the narration. Most slides with text contained body text, and the body text of a slide (*e.g.*, bullet points, paragraphs) was most often covered by the presenter when: the slide was sparse with text (as in TED talks, and humanities presentations), the presenter read off of their slides (as in some classes), or when the presenter dedicated a long amount of time to discussion such that they semantically covered slide content (as in applied sciences). When partially describing text content, presenters typically simply skipped elements on the slide (*e.g.*, not describing a bullet point, title, or a text label reference), but occasionally summarized the text, as in Figure 3B where the presenter summarizes a whole slide of text in a couple sentences.

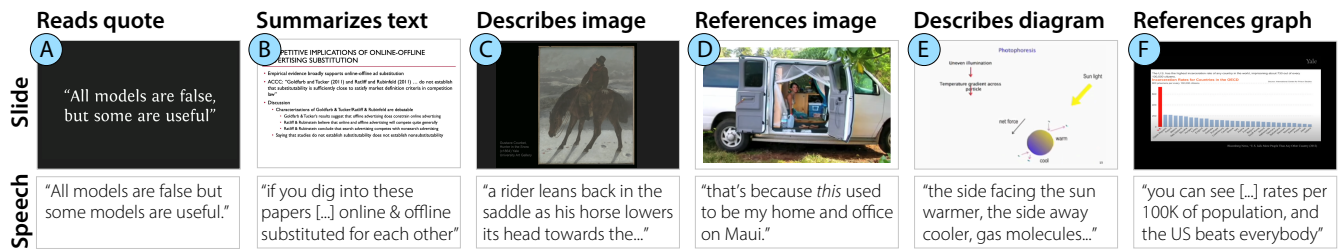

**Figure 3: Presentation slide examples with corresponding speech.** In (A) a quote is read nearly verbatim, but in (B) the speaker condenses half a slide of text into the phrase "online and offline substituted for each other". In (C) an art historian describes an image in vivid detail, but in (D) the van remains undescribed as the lecturer refers to the van as "this". In (E) a rare instance of a presenter describing not only diagram elements (sun), but also the relationship between the elements (faces away). In (F) a more typical example of a presenter telling the value but not explaining the visual context to understand its significance.

**Successfully describing media:** Presenters completely described the important visual content in their media most often in humanities presentations (Table 1). Humanities presentations featured fewer elements per slide, and spent more time on included images describing the visual content of images directly (Figure 3C). Diagrams and videos were the second most commonly completely described visual elements. For videos, presenters prepped the audience for fast videos, or narrated the video as it played; for diagrams, the diagrams were either simple extensions of the body text (*e.g.*, a set of text boxes connected sequentially), or the presenter carefully described the relations between the diagram elements (Figure 3E).

**Showing instead of telling:** Over 50% of the time, images on slides were either not described at all (none) or only briefly referenced in speech without a full description. In such cases, visual evidence or supporting figures were shown in parallel with the speaker's speech, or the speaker referenced the important element in the image without describing it (*e.g.*, "*this* used to be my home and office" without mentioning the van, Figure 3D). Surprisingly, media in TED videos was particularly unlikely to be described (6% complete for media, compared to 52% for media, generally). We found that the TED speaker guidelines actively discourage describing media: "the images represent what they're saying, so there is no need to verbally describe the images onscreen."[2]. A lack of reference or description appeared particularly often when making visual jokes, displaying sources meant to support the credibility of the talk (*e.g.*, a photo of an article that includes the source), or showing "awe-inspiring" content prompting audience clapping.

**Missing visual context:** When some but not all important concepts on a media object were described (little, half, most), the culprit was typically presenters instructing their audience members to look at important visual content without describing the surrounding visual context necessary to understand the reference. For instance, presenters described nodes of a diagram without describing the relationship between them, or described the main value on a graph or table without describing the trend, axes, or column labels (Figure 3F). Such descriptions would be accompanied with explicit references for the audience to observe the screen to gain visual context (*e.g.*, "you can see here"), and then use pronouns (it, this, these) and vague location reference (here, there) to continue referencing the visual content. In the case of text, some speakers would define a term without vocalizing the term itself.

**Dense slide content:** In cases of: (1) many elements on a slide, (2) dense slide elements such as tables or text media, or (3) a combination of visually dense elements and many other elements, the presenter would not practically be able to describe all of the visual content on their slides in the given amount of time. This phenomenon was particularly noticeable in the seminar category, as they displayed 0.56 more elements per slide than the aggregate sample, with a relatively poor completion rate (34%, Table 1). In seminar talks, speakers often spent a brief amount of time on slide with a complex informative graphic, or in-depth textual explanation (Figure 3B). They may gloss over the content itself, summarizing the key point, and then mentioning the graph (*e.g.*, "so that's what I'm showing there") before moving onto the next slide.

### 4.4 Reflection

Our analysis of slide presentations in the wild revealed opportunities to improve the description of visual content at the time of the presentation (VG1); for instance, by encouraging presenters to describe text and media using explicit description rather than implicit references (*e.g.*, nouns instead of pronouns, VG2), and to describe the relevant information surrounding an important point (*e.g.*, in a graph, table, or list) rather than only the important point itself (VG3). While real-time changes could improve descriptions in existing slides to some extend, editing the slide may be required to simplify potentially visually complex artifacts like diagrams and tables, and literature reviews such that they could be successfully described (VG5). As sparse slides were advantageous for presenters achieving high coverage of their visual elements (as in humanities, and TED with high text coverage and few elements per slide), and vice versa, encouraging authors to reduce their slide content could indirectly help presenters better describe the remaining content.

We note that although VG1 generally asks presenters to describe all slide content, we selected to analyze decorative elements separately from informative content (rather than omitting them completely) as decorative elements are likely to be less important based on prior guidelines in other domains [20]. Future work may refine VG1 for different domains and presentations to include fine-grained guidelines for content to describe (*e.g.*, omit descriptions of logos, or supplemental clip art).

---

[2]https://www.ted.com/participate/organize-a-local-tedx-event/tedx-organizer-guide/speakers-program/prepare-your-speaker/create-prepare-slides

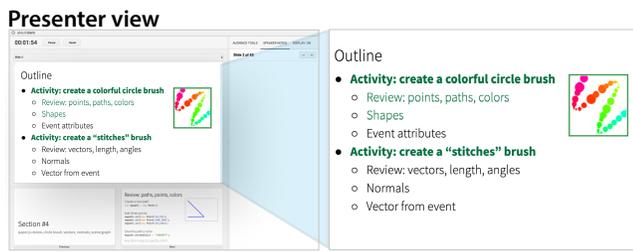

**Figure 4: The real-time feedback interface augments the *presenter view* with feedback, while the audience views the *main slide view* that shows the current slide without feedback. As the speaker talks, the presenter view shows words and images that they have covered, so that the presenter can glance to see what visual elements they have not yet described.**

## 5 SYSTEM DESIGN

To aid presenters in making their presentations more accessible, Presentation A11y instantiates accessible presentation guidelines in a system that provides feedback on presentation narration and their coupled visual aids through both *real-time feedback* and in *post-presentation feedback*. Presentation A11y's real-time and post-presentation interfaces give feedback related to the level of explicit correspondence between the presenter's verbal presentation and their visual aids, directly addressing guidelines (VG1-VG4). By encouraging presenters to create presentations that describe all of the visual content available to sighted audience members, Presentation A11y also encourages presenters to reduce visuals to the amount that they can describe (VG5), thus also dedicating more time per presentation element.

### 5.1 Interfaces

While our system design is applicable to many standard presentation interfaces including PowerPoint, Keynote, and Google Slides, we implement our interfaces by augmenting Google Slides (with a Chrome Extension) due to its broad availability and extensibility.

**Real-time feedback interface:** When giving a casual or unrehearsed slide-based presentation (*e.g.*, a lecture, informal work meeting, lab updates, reading groups), or when rehearsing for a presentation at a later date, the real-time interface provides presenters feedback on what elements they have and have not described to prompt further description as they present (Figure 4). The real-time interface extends the only Presenter View (i.e. the second-screen view seen only by presenters) such that the presenter can glance at their screen to see what slide elements need further description, but the slides shown to the audience remain unchanged. When the presenter describes text on the slide (*e.g.*, a vocabulary word and its definition), text on an image (*e.g.*, the x-axis title on a chart), or the contents of an image (*e.g.*, describes that a photograph contains two people looking at a computer), Presentation A11y highlights the corresponding slide element on the Presenter View (Figure 4, top). By default, Presentation A11y displays highlighted text and images using green, but presenters can select a color to best suit their needs and presentation background. Presentation A11y also reminds presenters to describe their videos by displaying an alert next to each video prior to playing it.

**Post-presentation feedback interface:** After a presentation, presenters can view their performance on accessibility feedback using the post-presentation feedback interface (Figure 5). The post-presentation feedback interface extends the slide authoring interface to provide overview feedback on the left-side slide overview panel, and detailed, element-specific feedback on the edit view of each slide. Using the post-presentation feedback interface, authors can preview performance to locate slides where they performed particularly well or poorly, then select a slide that might need further improvement (Figure 5A). On the slide view, authors can see (1) a summary of their slide performance on the overview pane that provides suggestions and coverage percentages, (2) element-level feedback on what elements they did and did not describe on the editable slide, and (3) the transcript of their slide performance.

The suggestions on the overview pane use the real-time interface results to determine what slide elements were described, then provide specific per-element template suggestions based on the element type (Figure 5B). The suggestion templates included "Remove the following text elements or add a description: ___" for unmentioned words on the slide or within text elements, and "Remove, describe, or add image alt-text for the following image elements: ___" for unmentioned images. Presenters can mouseover suggestions in order to highlight the corresponding location in the slide.

Presenters can interactively update their slides to be more accessible in the future by following interface suggestions: to delete undescribed elements from the interface, add description for undescribed elements to the script, or changing the automatically computed description of an image. As the user updates the slide, the interface interactively updates the slide summary and slide element feedback (*e.g.*, if the presenter deletes a graph that was not described, the displayed coverage percentage will increase).

### 5.2 Algorithms

In order to provide users real-time and post presentation feedback on how well their verbal narration covers the visual content on their slide, we employ a number of algorithmic methods (Figure 1). Our method for matching speaker speech to slide elements in real time builds on prior work [17, 42] that extracts slide segments and matches it to speech segments (*e.g.*, to help users navigate slides) using techniques such as word similarity and dynamic time warping. We design our algorithmic approach for the dual goals of: (1)

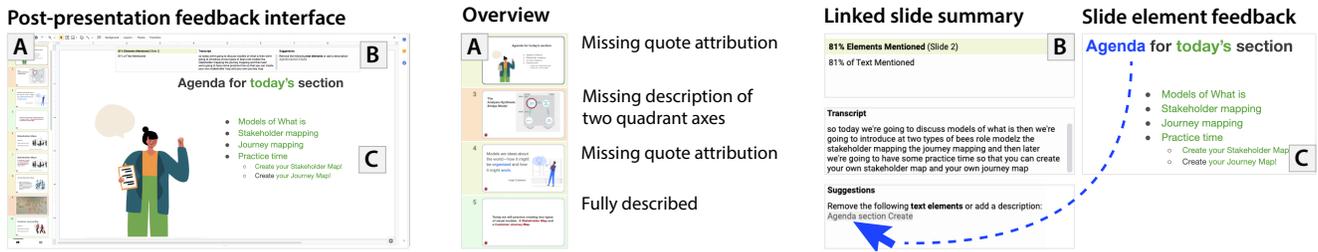

**Figure 5: The post-presentation feedback interface augments the *editor view* with feedback. As the speaker reviews their slides, the editor view shows the element-level coverage rate of words and images as well as the possible actions they could take to improve the slides accessiblity.**

supporting a live presenter by providing immediate feedback close to their locus of attention, and (2) encouraging presenters to further describe their slides through understandable and actionable feedback.

**Preprocessing:** Our system first automatically parses elements on the slide using the Document Object Model (DOM), and extracts their element type (text, image, or video). For images, we use Optical Character Recognition[3] (OCR) to recognize any text on the slide (*e.g.*, for an image picturing a street sign that reads "My way", we would extract "My way" along with bounding boxes for each word). Then, we assign a set of object labels to the image (*e.g. street sign, trees*) by first running Google Cloud Services Scene Recognition API (following prior work [42]), then adding additional labels for graph and chart elements; if an image is labelled "plot", "chart" or "graph", we add to its label set the set of other chart related words. We are over-inclusive rather than under-inclusive for image labels as we find that false negatives are more likely than false positives. For the purposes of matching spoken words to slide words, we consider all text words, OCR words, and image words as possible words that may match with a transcribed word. For matching slide words to incoming transcript words we use the stem of each word for matching such that "application", "applying", and "app" all match the root word "app."

**Predicting speech to slide alignment:** Our system runs real-time prediction whether or not the presenter turns the real-time feedback display on. For each presenter slide during the presentation, the real-time prediction method first predicts what slide elements are most likely to be said first by considering read order (distance from upper left-hand corner in the $x$ and $y$ directions). We use Google Cloud Services live transcription (up to 95% English accuracy [14]) such that we update the slide on each spoken word. When a speaker says a word, the system detects if the word is a stop word (*e.g.*, the, a, an [10]). If the spoken word is not a stop word, the system marks the first instance of that word with the highest probability as covered (and highlights the word or corresponding image in the interface if the display is on). For instance, if the first word the speaker says is "Review" and the word "Review" appears twice on the slide, our interface will highlight only the first instance (Figure 4). After the speaker says each word, our system increases the probabilities of all words subsequent to that word in the read order (*e.g.*, the next word to the right of the spoken word will have the highest probability). For example, if a presenter were to start by talking about the stitches brush (instead of the circle brush) the probabilities would update after the presenter said "stitches", such that if the presenter then said "Review" it would highlight the second instance (below stitches) rather than the first (Figure 4). For non-text images, we highlight the bounding box if any label word is mentioned. For images with text, we highlight the bounding box if a label word is mentioned, and highlight bounding boxes of individual OCR words when they are mentioned. For instance, if a user describes a graph "on your left, this graph shows..." the bounding box of the graph will highlight. As the user further describes elements within the image (*e.g.*, the graph title, axis labels) the predicted bounding box around each element will also highlight.

**Streaming transcripts:** From our transcription service we receive interim and final transcripts (rev.ai and otter.ai also provide interim transcripts). Interim transcripts indicate the most recent prediction, while final transcripts sometimes revise the transcribed segment to improve accuracy. To keep highlighting immediate, while avoiding duplicating words (*e.g.*, from multiple interim results) we concatenate the most recent interim result to all final transcripts for the current slide and re-compute spoken words per slide on each interim result received. To avoid flickering of highlights when transcription results oscillate between similar sounding words, we maintain a record of all words "ever spoken" such that we never un-highlight a word for the real-time interface.

## 6  PRESENTATION AUTHOR STUDY

To assess the effectiveness of Presentation A11y for improving presentation accessibility, we conducted a study with 16 presentation authors presenting their own slides. We explore two key questions:

- Does Presentation A11y's real-time feedback help presenters better cover their slide content?
- Does Presentation A11y's post-presentation feedback help presenters identify accessibility-related slide improvements?

### 6.1  Method

**Participants:** We recruited 16 presentation authors using departmental mailing lists that the authors had access to. Participants were 20-29 years old (12 female, 4 male) with occupations including

---

| ID | Venue | Topic | Script Rating | # Slides |
|----|-------|-------|---------------|----------|
| 1 | Class Project | Logic | 6 | 13 |
| 2 | Conference | Social Science | 5 | 56 |
| 3 | Conference | Haptics | 3 | 35 |
| 4 | Class Project | UI Design | 1 | 23 |
| 5 | Class Project | Drones | 3 | 17 |
| 6 | Lecture | Design Methods | 4 | 41 |
| 7 | Conference | Documentation | 5 | 33 |
| 8 | Lecture | Writing | 4 | 21 |
| 9 | Lecture | Design Methods | 5 | 22 |
| 10 | Class Project | Data Science | 3 | 13 |
| 11 | Class Project | Industrial Design | 5 | 22 |
| 12 | Conference | Education | 4 | 20 |
| 13 | Conference | Machine Learning | 2 | 26 |
| 14 | Lecture | Games | 2 | 13 |
| 15 | Conference | HRI | 7 | 21 |
| 16 | Class Project | Gender | 3 | 12 |

**Table 2: Participant IDs with the venue, topic, script rating ($\mu$ = 3.88, $\sigma$ = 0.93), and slide count ($\mu$ = 24.25, $\sigma$ = 11.5) for the presentations they gave during the study. The Script Rating represents how presenters rated the level of scripting for their presentations from 1 (not at all scripted) to 7 (fully scripted).**

students, teaching assistants, and designers. Presenters selected one of their own existing presentations to bring to the study (Table 2). The presentations represented a variety of venues (*e.g.*, conference, course project presentation) and topics (*e.g.*, writing style guide, machine learning tutorial). Three participants had prior experience creating accessible presentations. The participants included native and non-native English speakers.

**Materials and tutorial:** We preprocessed participants' slides by manually exporting any PowerPoint or Keynote slides to Google Slides, and checking that all image and video URLs were available to our system (*e.g.*, access was not restricted to a particular user). Participants installed our system's browser extension and selected: (1) a slide number of their presentation to serve as the half way point, and (2) a time estimate for how much time each half of their presentation would take (an equal amount of time for both halves). We provided participants with a reference sheet with a set of guidelines for non-visual presentation accessibility (*e.g.*, describing text and images on the slide, using nouns instead of pronouns, describing videos), and a hypothetical audience to target their in-study presentation to: their original audience where one or more audience members had visual impairments or would not be able to see the slides during the talk. To familiarize participants with the real-time feedback interface, we gave a 5-minute tutorial in which participants presented a set of practice slides.

**Procedure:** Our study consisted of: a *real-time feedback phase* in which participants presented their presentation, and a *post-presentation review phase* in which participants reviewed their presentation to identify accessibility improvements. During the real-time feedback phase, participants presented all of their slides in order, presenting half of the slides using Google Slides with real-time feedback from Presentation A11y and half using Google

Slides without any feedback. We randomized the order in which participants received feedback such that an equal number of presenters received Presentation A11y's real-time feedback on their first presentation half (8 participants) and second presentation half (the other 8 participants). To maintain the time pressure that is common to presentations, we timed each presentation half and provided participants with 2-minute and 1-minute warnings. During the post-presentation review phase, participants viewed all of their slides with and without the post-presentation feedback interface (in a counter-balanced order), and enumerated changes that they would make. Between the real-time feedback phase and the post-presentation review phase, participants completed a 10 minute questionnaire. We concluded with post-task questions about their experience.

**Measures and analysis:** We analyzed participants' level of slide description according to accessible presentation guidelines using automated (for text) and manual (for text and media) analyses. For automated analysis, we obtained the *% coverage of slide words* by first removing generic stop words (*e.g.*, on, a, the) from the slide, then dividing the number of slide words spoken (using our speech to slide alignment algorithm) by the remaining words on the slide. The automated analysis assesses coverage of exact slide words reliably (subject to small transcription errors), but it does not capture cases in which a presenter describes text elements in a different way than the system represents them (*e.g.*, synonyms, paraphrases). To manually analyze slide text and media element coverage, the paper author who conducted the studies first extracted audio recordings for each of the two presentation halves from the full user study recording. An annotator, who did not know the conditions assigned to each audio clip, then applied the established codes to all extracted audio recording/slide pairs (as in Section 3). We used the mean score for all text elements as the *text coverage score*, and the mean score for all non-decorative media elements as the *media coverage score*. In total, the user study contained 904 slide elements for analysis.

We also collected subjective feedback for both real-time interfaces with three following quantitative metrics in our questionnaire: accessibility of speaker's narrations for blind and visual impaired audiences, helpfulness that interface provided as the reminder for speakers to describe slide visuals, and the level of distraction interface caused. All metrics were collected using 7-point Likert scale (the higher, the stronger metric score). For the post-presentation feedback, we tallied users suggested changes in each condition. We analyzed post-study questionnaire by grouping interview notes into themes, and returning to the interviews to extract quotes.

**Hypotheses:** We consider four hypotheses:

**H1.** Users will cover a higher percentage of words in their presentation when using Presentation A11y compared to the traditional interface, because Presentation A11y's real-time feedback will prompt users to say additional words.

**H2.** Users will achieve a higher text coverage score when using Presentation A11y compared to the traditional interface, because Presentation A11y's frequent feedback will prompt users to describe more of their text elements.

**H3.** Users will achieve a higher media coverage score when using Presentation A11y compared to the traditional interface, because frequent media-coverage feedback prompts users to describe additional media elements.

**H4.** Users will generate a higher number of accessibility-related slide improvements when using Presentation A11y's post presentation feedback interface, because the interface provides element-level coverage feedback and content-specific suggestions.

## 6.2 Results

To assess order effects for each evaluation phase, we ran 2x2 repeated-measure ANOVAs (interface condition) for the four metrics we evaluated, including (M1) text coverage rate, (M2) text and (M3) media coverage score, and (M4) post-presentation edits. The results showed that there was no interaction effect for each metric ($p_{M1} = 0.63, p_{M2} = 0.85, p_{M3} = 0.27, p_{M4} = 0.38$). There was a main effect for interface condition ($p_{M1} = 0.018, p_{M2} = 0.048, p_{M3} = 0.035, p_{M4} = 0.0002$), but no main effect for order ($p_{M1} = 0.94, p_{M2} = 0.41, p_{M3} = 0.78, p_{M4} = 0.89$). All effect sizes (partial eta-squared) evaluated above are over 0.06 (medium). A follow-up marginal mean analysis shows that our proposed interface significantly improved each metric compared with the traditional interface ($p_{M1} = 0.016, p_{M2} = 0.045, p_{M3} = 0.038, p_{M4} = 0.0001$).

**Coverage of presentation text:** Presenters described a higher percentage of their slide words when using Presentation A11y ($\mu = 57\%$, $\sigma = 0.16$) compared to the traditional interface ($\mu = 46\%, \sigma = 0.14$) and a dependent t-test indicates that this difference is significant ($F(15) = 2.72, p < 0.05$) (**H1**). Presenters also achieved a higher text coverage score when using Presentation A11y's real-time feedback interface ($\mu = 4.24, \sigma = 0.45$) when compared to the traditional interface ($\mu = 4.03, \sigma = 0.44$) and a dependent t-test indicates that this difference is significant ($F(15) = 1.08, p < 0.05$) (**H2**).

**Coverage of presentation media:** For participant presentations with images, those presented using the real-time interface achieved higher scores for media coverage ($\mu = 3.07, \sigma = 0.64$) than those using the traditional interface ($\mu = 3.07, \sigma = 0.93$). But, many presenters did not have media in both halves of their presentations. To conduct a pair-wise comparison for media coverage, we excluded 7 incomparable participants (*e.g.*, no images to describe in one presentation half). Presenters achieved a higher media coverage score when using Presentation A11y's real-time feedback interface ($\mu = 3.61, \sigma = 0.53$) when compared to the traditional interface ($\mu = 2.88, \sigma = 0.56$), and a dependent t-test indicates this difference is significant ($p < 0.05$) (**H3**). Examining the distribution of media coverage scores, participants received the score of "1-None" less often when using Presentation A11y (6% of media elements scored "None") than when using the traditional interface (21% of media elements scored "None"). On the other hand, participants scored "2-Little" more often with Presentation A11y (26% with our system vs. 11% with the traditional interface) indicating that feedback could have prompted participants to provide some description ("Little") to usually undescribed ("None") elements.

**Post-presentation improvements:** Participants identified more changes to improve the accessibility of their presentations when using Presentation A11y's post-presentation feedback ($\mu = 2.25$; $\sigma = 0.93$) than they did when using the traditional interface ($\mu = 0.69; \sigma = 1.00$). A dependent t-test indicates this difference is significant ($F(15) = 5.42; p < 0.001$) (**H4**). While 15/16 participants (all except P2) reported at least one accessibility improvement with the Presentation A11y's post-presentation feedback, only 6/16 presenters could name any accessibility-related improvement with the traditional interface.

**Qualitative feedback:** All 16 participants reported that while presenting with Presentation A11y's real-time feedback (1) their awareness of accessibility increased, and that (2) they adjusted their speech to follow the feedback, covering more slide elements. P10 summarizes use of Presentation A11y's real-time feedback:

> "The real-time highlighting brought me a sense that my talk was listened to and evaluated by blind and visual impaired audiences, reminding me that I cannot assume the audience could access all the visuals on the slides. Every time I saw a word was marked, I gained some sense of achievement that I made the slide content more accessible; vise versa, I would also be more aware of the elements that I haven't mentioned yet, since I knew there was a 'veil' that hid the visual information from the audio signals and I think I should be the one to disclose it." — P10

Similar to P10, some presenters paid close attention to the feedback continuously: *"I feel that I was playing a game like Whac-A-Mole [...] where I want to turn everything green"* (P1), while others preferred to glance at the feedback occasionally: *"After talking for a period of time, I would check on the screen to see if I have covered all the crucial points on the slide. [...] I tend to start elaborating more on those uncovered elements."* (P6). While all presenters cited that they changed their description according to the real-time feedback, P2 used Presentation A11y to present a completed talk and found the feedback to be an additional hurdle to the prepared presentation: *"I already set certain flows and structures for my narrations"* so the real-time feedback interface made them *"pay additional attention on the mentioned or unmentioned visuals, increasing my mental efforts during the presentation"* (P2). Overall, all 16 participants suggested they would use the real-time feedback interface in the future, either for only practice presentations (4 participants) or for both practice and live presentations (12 participants).

All 16 participants reported that the post-presentation feedback improved their awareness of the accessibility of their presentation. Using Presentation A11y's post-presentation feedback, presenters identified accessibility improvements including: removing excess slide elements that they do not ever describe, adding descriptions of important but unmentioned slide elements to the speaker notes (*e.g.*, section title, subtitle), simplifying complex media and text elements (*e.g.*, reducing a large table to only relevant numbers), and adding descriptions to the speaker notes for over-abbreviated text elements (*e.g.*, defining an acronym). P9 summarized finding unexpected changes with the post-presentation feedback: *"this review interface can help me improve my presentation which I plan to give*

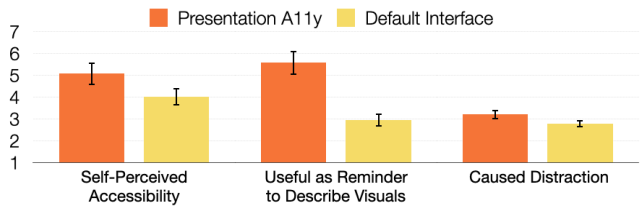

Figure 6: Participants rate Presentation A11y and the traditional interface (default interface) in terms of their self-perceived accessibility performance, the utility of the interface as a reminder to describe visuals, and whether or not the interface caused distraction. Ratings are on a Likert scale from (1-Low Agreement, to 7-High Agreement).

*during my TA session next week. [...] I found the same reference text on two of my slides [...] I also found there is more important text information on images than I expected that I should pay more attention to and think about how I would like to describe it in an accessible way next time"* (P9). All 16 participants stated they would use the post-presentation feedback interface when preparing slides in the future.

**Subjective ratings:** Participants rated their own presentations as more non-visually accessible (i.e. self-perceived accessibility) when using Presentation A11y's real-time interface than when using the traditional interface (5.06 vs. 4.00; $p < 0.05$) (Figure 6). Participants also found our real-time interface to be more helpful than the traditional interface for reminding them to describe the visual content on their slides (5.56 vs. 2.94; $p < 0.01$). However, participants rated Presentation A11y's real-time interface as more distracting than the traditional presentation interface although this difference was not significant (2.56 vs. 2.06; $p = 0.06$). The aforementioned statistical differences were verified by Wilcoxon sign-rank test.

**Observed errors:** During our study, 4.8% of words and 5.6% of images were not highlighted when a user mentioned the element (false negatives). These errors included: speech recognition errors (59.4%) and mismatches between speech and exact text representing the slide element (40.6%). False positives (elements mistakenly highlighted) did not occur. All participants cited occasional errors when asked. 12/16 said they ignored the errors (in real-time they saw the error and moved on), and 2 mentioned they did try to verbally cover those elements again.

## 7 DISCUSSION

Our work demonstrates the potential for providing real-time and post-presentation feedback to help make presentations more accessible. We include a reflection on limitations and future opportunities for research in crafting and authoring accessible media.

### 7.1 Accessible Authoring Tools

A limitation of our work is that the presentation feedback that it gives is itself currently inaccessible. While we suspect that blind and low vision presenters are less likely to rely on visual content that they do not describe, future work may nonetheless look to make Presentation A11y accessible non-visually. More broadly, visuals in presentation are core to how presentations operate across a wide variety of disciplines and settings. The centrality of visuals in presentation may usefully be challenged; however, the development of tools that allow visual presentations to be created by everyone (including those who are blind or low vision) in concert is an important future direction.

### 7.2 Production-Time Video Accessibility

Our work focuses on making presentations accessible by providing in-situ presenter feedback such that in-person and recorded versions of the talk will be more accessible to blind and visually impaired audience members. A benefit to this approach is that, unlike audio descriptions where a third party (with or without domain expertise [30]) provides a narrated description of visual content in a video, blind and visually impaired audience members get descriptions directly from the author (with domain expertise). Our approach — giving automated feedback on the correspondence (or lack thereof) between verbal and visual content — could be broadly applied to narrated video content to help video creators make their videos more accessible. For instance, in a cooking video, we could detect the correspondence between ingredients shown in the video and what ingredients have been described (using our approach and video accessibility metrics [26]), prompting people to add to the script a mention of the visual content (*e.g.*, an unmentioned pinch of salt). Fresco et al. [33] and Fryer et al. [11] consider innovative ways to manually integrate audio description in the artistic process of content creation. Our work indicates an opportunity to encourage creators to produce more inherently accessible content at scale.

### 7.3 Support for Physical Presentations

Our study was conducted remotely, but in the future, we will explore how our system could support presentations in contexts such as physical lectures and meetings. When the presenter can physically move and interact with the audience, our system may benefit from capturing additional environmental context (*e.g.*, a camera feed of the audience and presenter) and predicting additional events to provide better description recommendations. For instance, if a speaker takes a poll by asking the audience to raise their hands, our system could prompt the presenter to describe the result by analyzing the presenter speech. If the presenter conveys information using gestures, or if an audience member raises their hand without prompting, our system could prompt the presenter describe the scene by using a camera and pose recognition (*e.g.*, OpenPose [5]) for event detection.

### 7.4 Improving Feedback for Media Elements

While our system detected whether or not a presenter covered a text element (*e.g.*, title, text body) with high accuracy, detecting whether or not a presenter covered an image or media element is more challenging. In practice, we were able to predict *if* a user covered an image element in their speech, but not *how well*. This is because a high-quality description of an image or media element is often context-specific and contains much more detail than is provided by automatic image recognition. To improve our feedback on how well a user has covered an image or media element, we implemented

(after the user study) an alt-text authoring option in Google Slides where users can add their own image descriptions that replace the automated image description to provide a better comparison point for their verbal descriptions for scoring. The addition of alt-text to images can also benefit blind and visually impaired people who view the slides after the presentation. In the future, we will continue to explore new computer vision techniques to provide more descriptive image descriptions for automatic feedback (*e.g.*, DenseCap [15]).

## 8 CONCLUSION

Many slide presentations remain inaccessible to blind and low vision people because the visual content is often incompletely described. In this paper, we quantified this problem via a large-scale analysis of existing presentation videos across different domains and venues. We then introduced Presentation A11y, a system that provides real-time and post-presentation feedback helping presenters understand what visual content they have described. Participants in our study found the feedback helpful, as it enabled them to both describe more during their presentations and also identify useful areas to improve their presentations offline.

## ACKNOWLEDGMENTS

This work was funded by the National Science Foundation. We also thank our study participants and reviewers for their time and valuable feedback.